# Mapping Students' AI Literacy Framing and Learning through Reflective Journals

A Hingle [a,1], A Johri [b]

[a] George Mason University, Fairfax VA, USA, https://orcid.org/0000-0002-6178-1256
[b] George Mason University, Fairfax VA, USA, https://orcid.org/0000-0001-9018-7574

## ABSTRACT

This research paper presents a study of undergraduate technology students' self-reflective learning about artificial intelligence (AI). Research on AI literacy proposes that learners must develop five competencies associated with AI: awareness, knowledge, application, evaluation, and development. It is important to understand what, how, and why students learn about AI so formal instruction can better support their learning. We conducted a reflective journal study where students described their interactions with AI each week. Data was collected over six weeks and analyzed using an emergent interpretive process. We found that the participants were aware of AI, expressed opinions on their future use of AI skills, and conveyed conflicted feelings about developing deep AI expertise. They also described ethical concerns with AI use and saw themselves as intermediaries of knowledge for friends and family. We present the implications of this study and propose ideas for future work in this area.

---

[1] *A Hingle*
ahingle2@gmu.edu

# 1 INTRODUCTION

Given the extensive impact of artificial intelligence (AI) tools, techniques, and systems on everyday life, there is a growing call for advancing AI literacy among students and the larger population more generally (Long & Magerko, 2020; Tenório et al., 2023). However, developing these competencies is challenging given the complexity of AI as a domain and because learners bring a vast range of prior knowledge, skills, and experiences to their experiences with AI (Hornberger et al., 2023) (Heyder & Posegga, 2021). We conducted a journal study to examine how students learn about AI and how they envision their learning, formally and informally outside the classroom, will affect their current and future engagement with AI (Barron, 2004). The research questions that guided this work were: 1) how do undergraduate technology students' reflections on learning and interacting with AI align with established components of AI literacy from prior research (evaluate, use, create, and ethically navigate AI) and 2) how do students envision and value the role of AI in their learning and future? This paper builds upon the methodological foundation established in prior work (Hingle & Johri, 2024). We extend the scope of analysis in our prior work to assess students' progress toward different AI competencies through their reflections. By doing so, we hope to understand more about how students build literacy about AI both in and beyond the classroom.

# 2 PRIOR WORK: FRAMING AI LITERACY

Research on developing AI literacy is a growing area of scholarship and interventions in the classroom (Long & Magerko, 2020; Ng et al., 2021; Tenório et al., 2023). A recent systematic review outlines the significant efforts in AI literacy over the past five years (Almatrafi et al., 2024) and, based on an analysis of other articles, conceptualizes AI literacy as the ability to recognize, evaluate, use, create AI, and the capability to navigate AI applications ethically. To achieve momentum in improving AI literacy, the review recommends various instructional activities, including introducing everyday examples and supplementing that with knowledge of AI tools and applications and awareness. Finally, education about the limitations of AI, including ethical concerns such as fairness, accountability, transparency, ethics, and safety, and understanding AI's role in our world and its impact on society, both present and future, is considered an essential element, too.

Recognizing or becoming aware of AI is a fundamental first step in building AI literacy. However, implementing formal measures to raise AI awareness is challenging because people often interact with AI in unnoticed ways in their daily lives. Identifying AI's subtle role across society is crucial for developing an informed and critically engaged population that can interact with AI technologies responsibly. To this end, Koenig (2020) presents a study that engages writing students in reflecting on their daily interactions with algorithms through media journals. The analysis suggests that students became more aware of the underlying mechanisms by documenting and analyzing their engagements with platforms like *Facebook*, *Amazon*, and *Google*. Writing the journals encouraged students to examine patterns and question their role in maintaining them. The study suggests that while students initially understand algorithmic processes at a basic level, reflective journaling prompts them to develop a more nuanced, critical awareness. The journals also highlighted that not every learner engaging with AI wants to develop AI. Some try to better understand the systems they use daily and be informed users.

To be consistent with prior work, this study situates AI literacy in terms of learners' ability to *recognize*, *evaluate*, and *use* AI and their focus on ethical or social aspects. However, it explores the development of these constructs through learners' self-reflections. From a methodological perspective, this study uses journals to allow students to describe their everyday interactions with a focus on the components of AI literacy. The journals also enable participants to build on their experiences each week and build on connections between interactions they discussed across the weeks. The easily explainable or apparent engagements often emerge in the first or second journal, and students feel the need to do more in subsequent journals.

## 3 METHODS

### 3.1 Data Collection through Reflective Journals

Reflective activities are recognized across curricula, allowing learners to grapple with knowledge and lived experiences to create meaning, integrate thoughts, and build a deeper perspective (Brockbank & McGill, 1998; Bolger et al., 2003). They allow learners to consider decisions, questions, and alternatives and synthesize the knowledge (Koenig, 2020). Journal-based data collection is commonly used in educational research to capture students' ongoing learning practices (Arndt & Rose, 2023). Our approach is built on previous studies using reflection to understand self-regulated learning (Schmitz & Schmidt, 2011; Roth et al., 2016).

The study participants submitted reflections through a reflective journal entry each week for six consecutive weeks (early October to mid-November 2023) on their interactions with AI. Participants were asked to answer an open-ended prompt that encouraged them to reflect on any aspect of their engagements – what they did, how they interacted, where it occurred, and what impression the interaction left on them. The prompt given to students was:

> *In the previous week, document where and what you have noticed mentioned about AI: at your campus or outside of your campus? Did the description of AI leave you with a positive, negative, or neutral impression? Why or why not?*

The prompt was designed to be broad to capture various AI-related experiences and reflections while avoiding guiding students' thinking of AI in a specific. Broad categories of AI literacy were identified during the development of the reflective journal and the prompts from prior reviews (Long & Magerko, 2020; Tenório et al., 2023; Ng et al., 2021). These included both technical and non-technical competencies, in addition to implementations across different use cases.

### 3.2 Participants

In September 2023, twenty-two (22) students were recruited from a course designed to discuss the impact of technology worldwide to participate in this study. The participant's self-described gender ratio was 11 female: 11 male. The mean age of the participants was 23 years (range: 19-32 years). Most participants were full-time, pursuing a BS in Information Technology with various specializations, including development, cybersecurity, databases, and health information technology. The participants completed an average of 80 course credits (range: 40-121). Participants were given a $48 gift card to complete the six journal entries. No course credit was offered for participating.

## 3.3 Data Analysis Procedures

We used a hybrid approach to analyze the 22 collected reflective journal entries, which comprised six entries each for 132 reflective journal entries. First, we used emergent coding to capture all participant ideas due to the varied content of the journals, and second, we used the AI literacy framework (aware-knowledge-use-evaluate-create) to provide structure to the analysis. The approach was similar to Koenig's study on using journals for algorithmic literacy awareness (Koenig, 2020). After collecting the data, two researchers first read through the journals and highlighted interesting journal entries from which a set of initial codes was generated. Reviewing these initial codes, the researchers recognized similar ideas to those highlighted in earlier literature reviews - a focus on awareness, knowledge, usage, evaluation, and the ethical implications of AI use, in addition to developing AI. Using the five constructs, the researchers re-coded the data with the structure as a guide, ensuring that any interesting codes that did not fit in were retained. These codes were then grouped into themes, which were reviewed and finalized.

## 4 FINDINGS

Themes across the journal entries suggest that students were aware and largely cautious of AI's presence in their daily lives. They frequently connected their course learning with experiences outside the classroom. Many participants went beyond simple awareness and described their hopes and fears of AI and their expectations for what literacy could be, aligning with the more advanced elements of AI literacy. In this section, we present the findings from the reflective journal entries, using the categories of awareness, knowledge, application, navigating ethically, and development. Throughout, students describe their perceptions of the value of learning about and using AI and the potential impact they expect for their future.

### 4.1 Awareness of AI

Across the reflective journal entries, participants discussed their awareness of AI and a level of recognition or awareness they believed general people should possess. In at least one of the journal entries they wrote, all the participants described an expectation for a minimum level of societal understanding. Most participants described the lack of knowledge as posing as a personal and professional barrier.

> *As a student, I think learning about AI is not a choice. We have to be comfortable with AI if we want to do well in this field. But I am not sure if it is the same for other majors. I think they need to know what it is, but I don't think they really need to learn how to program or do machine learning. [P1 Week 3]*

Participant 1 highlighted that though the basic level of understanding will differ depending on the field of study, awareness is still important. Several participants, such as Participant 9, emphasized the understanding component to extend beyond learners in the classroom to a general awareness of AI for people:

> *I was a little concerned after [class discussion] because I have family that I think may be in danger because AI can do their work so much faster and cheaper. It is hard to talk to them about these issues because they don't know much about AI or technology at all. [P9 Week 5]*

Participant 9 highlighted the challenge of engaging in discussions with individuals who lack a fundamental awareness of AI concepts, let alone understanding the details, and that they may be disadvantaged in how they can adapt.

### 4.2 Knowledge of AI

While awareness refers to a general recognition of AI's existence and familiarity with where AI is being used, knowledge refers to AI concepts, techniques, and skills. Most participants distinguished between a high-level or general understanding of AI concepts and a more technical understanding of building and developing AI models. They noted that a macro-level understanding encompasses knowing what AI is, how it can be applied across various domains, and its potential societal impacts. This type of understanding is crucial for making informed decisions about AI's use and recognizing its broader implications. Participant 7 described wanting to learn some of the technical components but mostly being interested in how the systems work:

> *I thought [article about AI job changes] was interesting because it highlights how important AI will be going forward for everyone to have some understanding of. I personally don't intend to be a developer, but I still find the technical side of these systems to be very interesting. I think having deeper knowledge, even if I don't want to build AI, should give me an advantage over others. [P7 Week 3]*

Supporting this distinction, some students highlighted that though what they encountered was highly technical, they persevered as they found this one avenue to isolate the concepts they would likely need in the future. Participant 17 described coming across what seemed to them like an advanced topic but approached it anyway with an open mind to learning:

> *I noticed another seminar from the stats department titled [seminar title] about using AI models to understand complex problems around energy and the environment. The name was intimidating, but I still attended. I still have a lot to learn but I want to do this kind of work in the future so I am happy to listen. [P17 Week 5]*

Overall, participants distinguished between a higher-level macro-thinking and learning about AI compared to learning to create.

### 4.3 Using AI

As expected, many participants described the different AI uses they engaged with. These included software, systems, and tools as a service to accomplish a task. In addition to describing how they used AI, many participants articulated that using AI is highly related to their learning goals. They did so intending to enhance their competencies with tools they assumed would be useful in the future. They also often expressed a broad desire to stay knowledgeable on current tools and systems in a rapidly evolving field. Some participants, such as Participant 20, described their goals in learning specific skills that would be useful to them later in their careers:

> *In my [cybersecurity class], we talked about Splunk and the different AI capabilities that are available in the software and how they make things easier for administrators (like the anomaly detection features). Those kinds of hands-on classes feel especially useful and are why I wanted to take electives that talk about AI in them. [P20 Week 2]*

Similar to the discussions on a general understanding of AI, some participants described their operational intentions to understand and efficiently use the systems.

Participant 22 described how recruitment applicant tracking systems (ATS) serve as an example of applications everyone would likely need to be useful:

> I don't want to go into programming, but I think I will probably use some kind of AI in my work either way, so knowing how they work is important. One of my group members brought up how most companies use ATS software to screen resumes, and even there, knowing how it works helps you get through the process. [P22 Week 6]

Generative AI (GenAI) emerged in these discussions often, and there was variety in the breadth of anticipated use.

> I already use ChatGPT every day, and I have full intention to make the most out of the additional power it provides to a more than casual technology user like me. Getting better at prompting is something I plan to do sooner than later. [P8 Week 6]

### 4.4 Evaluate AI

Though often described as two separate constructs, participants frequently discussed ethical and societal implications in concert with critically evaluating AI. The participants primarily indicated unfamiliarity with technology ethics, with many encountering the topic for the first time in the course. Nonetheless, as Participant 22 highlights, they expressed a strong interest in further exploring the subject from other informed perspectives:

> I have started following people and groups on LinkedIn that talk about the ethics of using data in different ways. [P22 Week 6]

Participants persistently raised the issue of data bias, likely reflecting its coverage in their coursework. Notably, they demonstrated a deeper level of critical thinking on this issue. As Participant 9 described, data bias can affect the entire process from collection to any decisions made as a result:

> There are so many issues with the bias and types of data that are already in the system, I don't know if these can ever really be fixed. The bias in the data is built in from the start and it is used throughout the process. So, is there an easy way to even go about working with this type of data? I really don't think so. [P9 Week 2]

Participants also connected the implications of bias in data collection and its utilization in models, highlighting how these factors can preserve discriminatory processes and procedures.

> If not properly trained, AI can perpetuate and amplify existing biases in data, leading to discriminatory outcomes, especially against marginalized groups. [P2 Week 2]

Although the participants were primed for these discussions through course readings, their choice to address these topics in a journal, where they had the freedom to discuss anything, is a positive finding for transferring learning beyond the course material. Participants gravitated towards topics on AI's social impact, such as bias or ethics. In the analysis, participants often began describing other concerns associated with data-driven decision-making, such as those of power dynamics, surveillance, and safety, but could not specifically describe them.

## 4.5 Developing AI

While including the development of AI or the use of machine learning techniques specifically in AI literacy measures is contested (Carolus et al., 2023), participants often described their intentions for learning about AI through this framing. This may be because all the participants were in a technical program where they were required to take programming, algorithms, and machine learning courses. Most participants described the feasibility of applying these skills in the future. They described the knowledge and behaviors they acquired as highly relevant and transferable to academic, professional, and entrepreneurial contexts. Some participants also thought of using GenAI similarly:

*It was really interesting to read about how [Copilot] is being included in the GitHub environment and what this could potentially mean about generating code or helping people write code. I think this is a worthwhile idea to work on because coding can be difficult, and we want as many people to understand how to code. I find writing code to be intimidating, which is why I don't plan on becoming a programmer, so having help like this makes me feel like I can use it when I want. [P11 Week 5]*

Finally, some participants expressed a desire to enhance their AI literacy to a level sufficient for pursuing entrepreneurial ventures and building a business:

*But with AI being so good at looking for patterns, maybe it is better that [agricultural systems, such as those used to efficiently water crops] are taken care of by AI. I am really interested in how these systems work, and after working for a few years, I want to be able to create my own business that uses AI. [P20 Week 5]*

## 5 DISCUSSION AND IMPLICATIONS

The findings present several considerations regarding both research questions. First, through the framing of AI awareness or knowledge, the idea of a baseline AI understanding was universally described by all the participants. This aligns with the discussions around AI literacy that have emerged over the last decade (Druga et al., 2019; Long & Magerko, 2020; Touretzky et al., 2019). Some participants argued that this would ensure everyone could make the most of technological advances, but others took a protective stance regarding livelihoods and consumer protection. The participants presented a case for a technologically aware population at a general level as a minimum. However, defining what the minimum should be is challenging. One major challenge is that learning about AI basics can be difficult if the learner does not understand the prerequisites. Especially when participants described general AI literacy, when taking the perspective of a friend or family member, they brought to attention that they were completely unaware of AI. The disconnect between how common AI tools appear in everyday spaces was lost to them. It required the participants to step in as intermediaries, enforcing their understanding and concerns about AI with others.

Access to resources, both the ability to pay for an AI subscription, education on the topic, or mentorship may exacerbate the issue. Celik presents the digital divide (unequal access to technologies) and computational thinking as determinants of AI literacy (Celik, 2023), and this is reflected in some responses in this study. Especially when participants described a generalized level of literacy, when taking the perspective of a friend or a family member, they brought to attention that they were completely unaware of AI. The disconnect between how common AI tools show up in everyday spaces was lost to them and required the participants to step in as intermediaries, a role that provides them with an opportunity to learn by teaching (Palinscar & Brown, 1984). Still, it is a role that they may be unprepared for.

Regarding using AI, participants highlighted how their interests played a role in naturally building literacy. Participants predominantly considered learning about AI as a hobby, which could be part of the culture of being in a technology-focused program. Interest in computing topics can play a role in defining the learner's goals (Bollin et al., 2020). However, this may be different if these participants were from non-technical programs. As such, extending career mentoring, coaching, and realistic expectation setting should be a part of building AI literacy among students. In this study, most participants implied that they did not have mentors in this field.

Most participants expressed interest in using the technical skills they were developing around AI in their future careers. However, they were split on how to do so. This mirrored the discussion on baseline understanding and the learners' goals. Participants articulated different personal decision-making layers and thought about how and why they are learning about AI. Access to information, certifications, and training are abundant, and focusing on improving one's competencies with AI involves an opportunity cost where the student could be learning or doing something else. Mentorship may also have affected this intentionality.

From the reflections, it was apparent that the participants were navigating the hype of AI while being cautious of the societal implications. Some participants implied technosolutionist thinking of using AI to solve every problem. However, most participants expressed uncertainty about how the skills they were developing would be used, even though they still thought they would be useful. Others described purposely taking on challenging topics in seminars and courses despite not knowing how useful they would be because they thought this was their expectation as students. For many participants, it appeared that the debate regarding the importance of AI had already been resolved – AI was seen as ubiquitous and critical – and they were merely responding to this established consensus.

## 6 LIMITATIONS AND FUTURE WORK

The student population volunteered to participate and thus self-selected, and all majored in information technology, which limits the generalization of this study. The emergent analysis presents some findings in defining themes, but this study did not evaluate the correlation between the quantitative items. We encourage further exploration with explainability as a central focus. In this work, we do not explore any data from follow-up discussions. However, future work will explore the impact of a post-intervention discussion, either at an individual or group level. This may be exceptionally useful for building AI literacy and awareness. Finally, there are limitations to this work as AI is a fast changing field and new developments are likely to change students' viewpoints and understanding frequently.

## 7 ACKNOWLEDGEMENTS

This work is partly supported by US NSF Awards 2319137, 1954556, and a USDA/NIFA Award 2021-67021-35329. A research grant from George Mason University also supports this work. Any opinions, findings, conclusions, or recommendations expressed in this material are those of the authors and do not necessarily reflect the views of the funding agencies.